\RequirePackage{fix-cm}
\documentclass[smallextended]{svjour3}       
\smartqed  
\usepackage{graphicx}
\usepackage[utf8]{inputenc} 
\usepackage{amsmath,amsfonts}
\usepackage{arydshln}
\usepackage[inline]{enumitem}
\usepackage{xspace}
\usepackage{booktabs}
\usepackage{listings}
\usepackage{url}
\usepackage{enumitem}
\usepackage{flushend}
\usepackage{xcolor}
\usepackage{multirow}
\usepackage{multicol}
\usepackage{cite}
\usepackage{hyperref}
\begin{document}

\title{Secure Software Engineering in the Financial Services: A Practitioners' Perspective}
\titlerunning{Secure SE in the Financial Services}        

\author{Vivek Arora         \and
        Enrique~Larios Vargas \and
        Maurício~Aniche \and
        Arie van Deursen 
}


\institute{
   Vivek Arora \at Delft University of Technology, Delft, The Netherlands \\
   \email{v.arora@tudelft.nl}  \and
   Enrique Larios-Vargas \at Delft University of Technology, Delft, The Netherlands \\
   \email{e.lariosvargas@tudelft.nl}   \and
   Maurício Aniche \at Delft University of Technology, Delft, The Netherlands \\
   \email{m.f.aniche@tudelft.nl}   \and 
   Arie van Deursen \at Delft University of Technology, Delft, The Netherlands \\
   \email{arie.vandeursen@tudelft.nl}   
}

\date{Received: date / Accepted: date}

\maketitle

\begin{abstract}

Secure software engineering is a fundamental activity in modern software development. However, while the field of security research has been advancing quite fast, in practice, there is still a vast knowledge gap between the security experts and the software development teams. After all, we cannot expect developers and other software practitioners to be security experts. Understanding how software development teams incorporate security in their processes and the challenges they face is a step towards reducing this gap. In this paper, we study how financial services companies ensure the security of their software systems. To that aim, we performed a qualitative study based on semi-structured interviews with 16 software practitioners from 11 different financial companies in three continents. Our results shed light on the security considerations that practitioners take during the different phases of their software development processes, the different security practices that software teams make use of to ensure the security of their software systems, the improvements that practitioners perceive as important in existing state-of-the-practice security tools, the different knowledge-sharing and learning practices that developers use to learn more about software security, and the challenges that software practitioners currently face when it comes to secure their systems.

\keywords{software security \and secure software development \and security tools \and empirical software engineering \and qualitative research.}

\end{abstract}


\newcounter{practice}
\newcommand{\practice}{\refstepcounter{practice}\textbf{(SP\thepractice)}}

\newcounter{observation}
\newcommand{\observation}{\refstepcounter{observation}\textbf{(S\theobservation)}}

\newcounter{benefit}
\newcommand{\benefit}{\refstepcounter{benefit}\textbf{(B\thebenefit)}}

\newcounter{limitation}
\newcommand{\limitation}{\refstepcounter{limitation}\textbf{(L\thelimitation)}}

\newcounter{challenge}
\newcommand{\challenge}{\refstepcounter{challenge}\textbf{(Ch\thechallenge)}}

\newcounter{resource}
\newcommand{\resource}{\refstepcounter{resource}\textbf{(R\theresource)}}

\newcommand{\factor}[1]{\textit{#1}}
\newcommand{\source}[2]{\textit{#1~\textsuperscript{(\em #2)}}}
\newcommand{\block}[2]{\textit{#1~\textsuperscript{(#2)}}}

\newcommand{\fix}[1]{\textcolor{red}{#1}}
\newcommand{\added}[1]{\textcolor{blue}{#1}}

\newcommand{\cooltitle}[1]{\vspace{1mm}\noindent\textbf{#1}}

\newcommand{\numberoffactors}{26\xspace}
\newcommand{\numberofsurveys}{115\xspace}
\newcommand{\numberofinterviews}{16\xspace}
\newcommand{\numberofnewfactors}{8\xspace}
\newcommand{\numberofcompanies}{11\xspace}

\newcommand{\hist}[1]{(\includegraphics[height=2.5mm]{#1})}
\newcommand{\histt}[1]{\includegraphics[height=2.5mm]{#1}}

\newlength\replength
\newcommand\repfrac{.33}
\newcommand\dashfrac[1]{\renewcommand\repfrac{#1}}
\setlength\replength{1.5pt}
\newcommand\rulewidth{.6pt}
\newcommand\tdashfill[1][\repfrac]{\cleaders\hbox to \replength{%
  \smash{\rule[\arraystretch\ht\strutbox]{\repfrac\replength}{\rulewidth}}}\hfill}
\newcommand\tabdashline{%
  \makebox[0pt][r]{\makebox[\tabcolsep]{\tdashfill\hfil}}\tdashfill\hfil%
  \makebox[0pt][l]{\makebox[\tabcolsep]{\tdashfill\hfil}}%
  \\[-\arraystretch\dimexpr\ht\strutbox+\dp\strutbox\relax]%
}
\newcommand\tdotfill[1][\repfrac]{\cleaders\hbox to \replength{%
  \smash{\raisebox{\arraystretch\dimexpr\ht\strutbox-.1ex\relax}{.}}}\hfill}
\newcommand\tabdotline{%
  \makebox[0pt][r]{\makebox[\tabcolsep]{\tdotfill\hfil}}\tdotfill\hfil%
  \makebox[0pt][l]{\makebox[\tabcolsep]{\tdotfill\hfil}}%
  \\[-\arraystretch\dimexpr\ht\strutbox+\dp\strutbox\relax]%
}

\newcommand{\mauricio}[1] {\textcolor{blue}{\textbf{[Maurício: #1]}}}

\newcommand{\vivek}[1] {\textcolor{cyan}{\textbf{(Vivek: #1)}}}

\section{Introduction}
\label{label:introduction}

The advancements in software security mechanisms have not yet been able to stop the increase in the number of reported security vulnerabilities and attacks on critical software systems. The NIST's NVD (National Vulnerability Database) statistical figures report more than 100\% increase in reported vulnerabilities since 2016~\cite{NVDStati0:online}. According to the 10\textsuperscript{th} Veracode report on the state of software security 2019, two in three applications fail to pass the compliance check based on OWASP~\cite{OWASPTop10:online} and SANS~\cite{Top25Sof45:online} standards. The SANS/CWE's report on the ``Top 25 most dangerous software errors''~\cite{Top25Sof45:online} shows that memory buffer errors, cross-site scripting, and information exposure are still prevalent security risks. 

It is easy to blame software developers and development teams for the lack of security in software systems. After all, security experts have been quite rapidly advancing the field when it comes to API security guidelines, tools, and secure coding practices. Meanwhile, it is not uncommon to see development teams following poor secure development practices or having lack of security know-how~\cite{Conklin2007,Wurster2018}. However, the community now recognizes that it is unrealistic to expect developers to become security experts~\cite{7676144,Wurster2018,Acar2017}. There is simply a considerable knowledge gap between the skills required for developing a software system and to ensure that it is secure. We therefore argue that a critical question is how to support software developers in adopting effective security practices and knowledge during the software development process.


Historically, Saltzer et al.~\cite{Saltzer1975} first mentioned the requirement of a psychological base while discussing security design principles. Zurko et al. ~\cite{10.1145/304851.304859} later introduced the term ``user-centered security'' to refer to the security models, mechanisms, and systems that take usability (i.e., how end users, such as developers or system administrations, interact and make use of them) as a primary goal. This new research area started to include the human and social aspects into security research. Since then, we see researchers considering security problems that were once only explored from the technical perspective, being now also explored from the human side. For example, Whitten et al.~\cite{10.5555/1251421.1251435} emphasized that email security is a usability problem rather than a technical one. 
Since then, usable security research has produced many interesting results (e.g., ~\cite{10.1145/2501604.2501609,10.5555/1251421.1251435,sheng2006a,garfinkel2005a,ruoti2016a,biddle2012a,komanduri2011a}). 

However, seeing other software practitioners (e.g., developers, security engineers, product owners) as also the consumers of the security knowledge generated by the security experts is somewhat less frequent.
Acar et al.~\cite{Acar2017} realized this and included developers in the context of usable security in their research. They utilized the learning from two decades of research in usable security for end-users, and proposed a developer-centric research agenda. The agenda consisted of developing new methodologies and security mechanisms, as well as understanding the developers' motivation, attitudes, and security knowledge. 

Inspired by their observations, we aim at understanding how software development teams have been ensuring the security of their software systems. More specifically, we seek to discern how security is incorporated in their development processes, what practices and tools they use, what types of challenges they face, and how they spread security knowledge. In particular, we focus on software teams that build financial services, as security is paramount in such systems. We argue that studying the development processes of such sensitive applications can provide researchers and practitioners with extensive insights into the real-life secure software engineering practices.

To that aim, we performed a qualitative study, based on semi-structured interviews of a cohort of 16 software practitioners involved in the development of applications for financial services from 11 different companies geographically distributed in three different continents. Our results make the following contributions:
\begin{enumerate}
    \item The security considerations that practitioners take during the different phases of their software development processes,
    \item A set of 23 security practices that software teams use to ensure the security of their software systems,
    \item A list of improvements that practitioners perceive as important in existing state-of-the-practice security tools,
    \item Different knowledge-sharing and learning practices that developers use to learn more about software security, and
    \item The challenges that software practitioners currently face when securing their systems.
\end{enumerate}

\section{Background and Related Work} 
\label{label:background}

Software security introduces different engineering aspects into the existing software development process. In the following sections, we discuss related work focused on the secure software development process, security tools, and the human factors in secure software engineering.

\subsection{Secure Software Development}

A variety of research in secure software development suggests improvements in the development process or creating proprietary development methods~\cite{Microsof69:online,bostrom2006extending,FIRDAUS2014546}. With the evolution of software development methods, we see the focus of researchers shifting towards the comparison of secure development methods and integrating security into the development process or vice-versa~\cite{Rindell2015,Matharu2015,Rindell2018,Ayalew2013}. 

Earlier work by Flechais et al.~\cite{Ivan2003} proposes a secure software engineering method that integrates risk and threat assessment with other project-specific parameters to provide security analysis during the development process. Later, Davis et al.~\cite{1306968} looked at the problem of developing secure software and identified that producing secure software has challenges at three levels: software engineering practices, technical practices, and management practices. To build more secure and reliable software, the authors provide short-term, mid-term, and long-term recommendations. The short-term recommendation focuses on improving the software development process with a very low defect rate ($<0.1$ defects/1000 LOC), mid-term on improving the security of the products and processes, keeping the focus on software specifications and design, and finally, the long-term recommendation focuses on preparation for security by certification, training and evaluating emerging technologies. While the authors provide a large set of recommendations and organizational changes, they agree that the recommendations are not realistic for most organizations.

With the adoption of the agile methodology by industry, a variety of work on agile security assurance explored the large gap between security assurance methods and agile methods~\cite{10.1145/1065907.1066034,6046004}. Beznosov et al.~\cite{10.1145/1065907.1066034} in their research, examined if the security assurance techniques and methods fit with the agile methods. They found that approximately half of the conventional security assurance methods (e.g., formal/information validation, external review, security evaluation, etc.) do not fit the agile methods. 
On the other hand, incorporating security practices in agile processes increases the developers' workload, resulting in dissatisfaction~\cite{6702438,10.1145/1065907.1066034}.

Related work by Assal et al.~\cite{Assal2018} explores how security fits in the software development workflow. Researchers identified that the real-life practices of the developers vary significantly. Researchers also identify the best practices from various resources and find that the developers generally ignore best practices. The authors pointed out that the best practices sometimes conflict with their role. For example, the developer carrying out the functional testing will be in conflict if security testing is blended into the task. Similarly, developers that are less (or not) equipped with security knowledge will not do justice with the assigned security task. Additionally, the company's culture, resource availability and security-incidents directly impact the adoption of security practices. In another study, Assal et al.~\cite{Assal2019}, explored how developers influence and get influenced by organizational processes. They found that the organization process needs better structure to support the developers. While the participants in their research were self-motivated but lack of support from the organizational processes is a roadblock in ensuring security. This study extends the results in secure software engineering by studying the security consideration and challenges in the development process of financial services where security is of utmost importance.

\subsection{Security Tools}

There is a good amount of research on static and dynamic analysis tools. For example, Astree\cite{Astree2007}, CPPcheck\cite{cppcheck2013}, Joern\footnote{https://joern.io/} are the static analysis tools for C language projects, Findbugs\footnote{http://findbugs.sourceforge.net/} for Java bytecode, and Coverity\footnote{https://www.synopsys.com/software-integrity/security-testing/static-analysis-sast.html}, SonarQube\footnote{https://www.sonarqube.org/}, Klocwork\footnote{https://www.perforce.com/products/klocwork} are multi-language commercial static analysis tools. Companies highly depend on automated tools to detect security vulnerabilities, mainly because with a large codebase, it is expensive to perform the manual code review~\cite{Beller2016,Piskachev2019}. The security tools can detect various security issues with the help of static and dynamic analysis techniques. They are crucial when finding bugs automatically without manual inspection~\cite{ayewah2008using}, such as SQL injections, issues with the pointer de-referencing, and off-by-one errors. Despite the benefits, developers find security tools hard to use~\cite{baca2009static,Sadowski2015} because of their high false-positives rate and their (lack of) usability aspects~\cite{johnson2013don,harman2018start}. Interestingly, Witschey et al.~\cite{witschey2015quantifying} in their research, identified that the security tools' adoption is correlated with the developer's experience and inquisitiveness. Researchers pointed out that adoption of security tools are more likely to happen if they is observed to be used by peers~\cite{witschey2015quantifying}.

T\"{u}rpe et al.~\cite{10.1145/2413296.2413300} in their research argue for the development of better security tooling in the context of interaction of threat actors. According to the researchers, the existing security tools are more focused on low-level identification of security issues. Additionally, the author shared a ``property degree'' (low to high) framework to visualize software security in multiple ways. A low or microscopic property degree corresponds to attack action, a medium or mesoscopic property degree corresponds to an attack object. Finally, a high or macroscopic property degree corresponds to the population of adversaries.

Sadowski et al.~\cite{Sadowski2015} point out the integration issues of static analysis tools with the developer workflow and with each other. They propose a program analysis platform, ``Tricorder'', and the philosophy on how they created the platform, which developers use across Google. They propose the guiding principles while creating future tools, such as, data-driven usability improvements (e.g., developers will decide if the analysis tool has an impact, empowering users to contribute, focus on workflow integration, etc).

In this research, we aim to add to the results by studying the aspects of security tools where software practitioners want to see the improvement to help them ensure security.

\subsection{Human Factors}

The security aspects from the end-user perspective have been studied a lot in the last two decades~\cite{Acar2017}, and recently the security focus has included developers and security experts. 

A recent systematic literature review focusing on the developers' context in secure software development by Tahaei et al.~\cite{Tahaei2019} identified multiple themes and research gaps. 
The identified themes include organization and context, structuring software development, privacy and data, third-party updates, security tool adoption, application programming interfaces (APIs), programming languages, and testing assumptions. Researchers highlighted gaps in experiment design by researchers, addressing security tools limitations, security issue communication by developers, learning support for developers, and security and privacy awareness among developers.

Acar et al.~\cite{Acar2017} produced a research agenda focusing on developers. They used the learning from two decades of usable security for end-users to produce a list of research questions aiming at developing new methodologies, developer's motivation, attitude, knowledge, and security mechanisms (API security, documentation, and tools). Pieczul et al.~\cite{Pieczul2017} explored the real-life challenges with developer-focused security, including the agenda by Acar et al.~\cite{Acar2017} and pointed out the need to look at it from a different perspective. The authors pointed out that treating developers as end-users can be a good first step; however, the software development process involves many individuals with different roles, responsibilities and limitations. Furthermore, the emphasis on end-user perspective, i.e., design for use before use, may not translate well with a developer-centered approach because the development task is more complex than secure usage of the artifact.

Thomas et al.~\cite{Thomas2018} found that application security is mainly performed by security experts rather than developers. They observed that separating security and development delays security bug fixing. Researchers emphasize the organizational level efforts and tools for software practitioners to reduce the workload.

In another study by Votipka et al.~\cite{votipka2020understanding}, the quantitative analysis of a developer contest on secure-programming showed that developers are more prone to make security errors because of the misunderstanding of security concepts. 

Related research work in the direction of learning support for software practitioners includes interviews and surveys. Acar et al.~\cite{Acar2016,Acar2017a}, highlight the information resources that developers trust and the impact of insecurities associated with them. The researchers looked into the 19 general advice resources (CERN, OWASP, etc.), which the developer looks towards when finding answers to security-related issues. They argue that the developers recognize the need to integrate security and upgrade their skills; however, the current ecosystem needs to address their needs and better evaluate existing knowledge acquisition resources. They report that the developers lean towards more accessible but less secure resources (such as StackOverflow\footnote{https://stackoverflow.com/}) because more robust and secure API documentation is hard to comprehend. It is highly likely that developers will not stop using such resources soon. The researcher interviewed 19 developers and a survey with 228 developers about privacy and security behavior. They found that there is a lack of education and interest in security and privacy. 

Additionally, researchers found that developers seek consultation online and among the social networks and friends~\cite{6876252}. Unfortunately, the online resources are found to harm code security~\cite{acar2016you}. Tahaei et al.~\cite{Tahaei2019} in their research, found that learning support for developers in secure software development is an area yet to be addressed well.

In this study, we take the viewpoint of software practitioners employed in different financial service companies worldwide. We aim to add to these results by consolidating the software practitioners' real-life security practices.

\section{Research Method}
\label{label:methodology}

The goal of this study is to \emph{understand how software development teams address security concerns in the context of financial services}. To that aim, we propose the following research questions:

\newcommand{\rqone}{\textbf{RQ$_1$}: How is security incorporated into the different phases of the software development life cycle (SDLC)?}
\newcommand{\rqtwo}{\textbf{RQ$_2$}: What practices and techniques do software practitioners apply to ensure security? }
\newcommand{\rqthree}{\textbf{RQ$_3$}: What are the limitations of security tools used by software practitioners?}
\newcommand{\rqfour}{\textbf{RQ$_4$}: What resources do practitioners use to learn and share knowledge about software security?}
\newcommand{\rqfive}{\textbf{RQ$_5$}: What challenges do practitioners face while incorporating security to the software development process?}

\begin{itemize}
    \item \rqone
    \item \rqtwo
    \item \rqthree
    \item \rqfour
    \item \rqfive
\end{itemize}

\begin{figure}
\includegraphics[width=\textwidth]{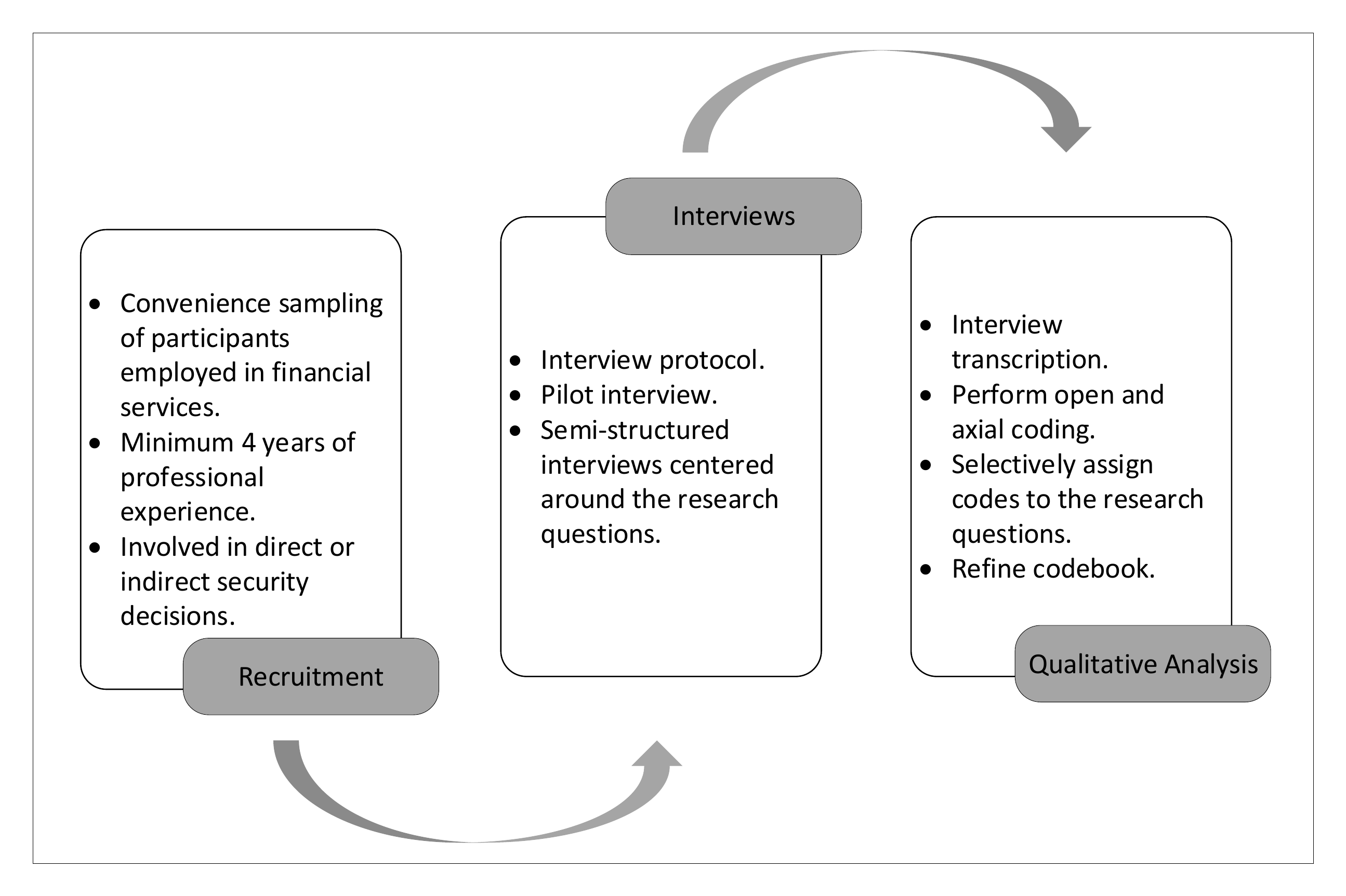}
\caption{Overview of the research methodology.}
\label{fig:methodology}
\end{figure}

We designed a qualitative study, based on semi-structured interviews with software practitioners involved in the development of applications in various financial services. Figure~\ref{fig:methodology} shows an overview of our research methodology. 
In the following sections, we describe the recruitment process, the participants' profiles, the interview procedure, and the data analysis.

\subsection{Recruitment and Participants}

We recruited participants that are currently employed in financial services companies and are involved in ensuring the security of the company's software systems. The pool of participants came from convenience sampling. The authors of this study invited professionals from their networks. Our selection criteria admitted only participants who had at least 4 years of experience in the software industry, and have been involved directly or indirectly in security decisions. We also aimed for a high diversity of financial companies. 

In the end, we recruited a cohort of \numberofinterviews participants from \numberofcompanies companies, which we detail in Table~\ref{tab:participants}. We observe that their professional working experience ranges from 4 to 32 years, companies range from small up to large enterprises and are based on different regions such as Europe (5 companies), North America (1 company), and South America (5 companies). 

\begin{table*}
\small
\centering

\caption{Profile of our participants (N=\numberofinterviews).}
\begin{tabular}{lp{4cm}rlllr}

\toprule
\textbf{} & \textbf{Role/} & \textbf{Years} & \textbf{}                              & \textbf{Company}                                 & \textbf{} & \textbf{Interview} \\
\textbf{ID} & \textbf{Function} & \textbf{of Exp} & \textbf{Company}                              & \textbf{Size}                                 & \textbf{Region} & \textbf{Duration} \\

\midrule
P1          & Cyber Specialist Leader                        & 9                                 & C1      & Large        & NA      & 56\\
P3          & Software Architect                             & 10                                & C1      & Large        & SA        & 70 \\
\hdashline

P2          & Cyber Specialist Leader                        & 13                                & C2      & Large        & SA        & 88\\
\hdashline
P4          & Cyber security manager                         & 11                                & C3      & Large        & SA        & 88 \\
\hdashline
P5          & Software Developer                             & 9                                 & C4      & Medium          & SA      & 111 \\
\hdashline
P6          & DevSecOps Engineer                             & 7                                 & C5      & Large        & EU & 60 \\
P7          & Developer                                      & 20                                & C5      & Large        & EU & 67 \\
P8          & Sr. Information Security Expert                & 7                                 & C5      & Large        & EU & 47 \\
P11         & Information Security Architect/Project Manager & 16                                & C5      & Large        & EU & 83 \\
\hdashline
P9          & Software Engineer                              & 13                                & C6      & Large        & EU     & 55 \\
\hdashline
P10         & Development Coordinator                        & 10                                & C7      & Large        & SA      & 84 \\
\hdashline
P12         & Chief Security Architect                       & 32                                & C8      & Small        & EU & 47 \\
\hdashline
P13         & Developer                                      & 8                                 & C9      & Large        & EU & 83 \\
\hdashline
P14         & Product Owner \& Data Scientist                & 4                                 & C10     & Large        & EU & 61 \\
P16         & Lead, Architect, Cyber Innovation           & 24                                 & C10     & Large        & EU       & 79 \\
\hdashline
P15         & Cybersecurity | DevSecOps Specialist           & 7                                 & C11     & Large        & SA       & 46 \\

\bottomrule

\end{tabular}
\vspace{0.02in}

Company size is based on their number of employees (small $<=$ 50, 50 $<$ medium $<=$ 250, large $>$ 250). The interview duration is represented in minutes. Regions are SA=South America, NA=North America, EU=Europe. Participants are grouped by their companies.

\label{tab:participants}

\end{table*}

\subsection{Interview Procedure and Analysis}

We conducted semi-structured interviews aiming at collecting the participants' descriptions, experiences, observations, and assessments of how software security is ensured in financial services. We opt for semi-structured interviews as they tends to encourage participants to share their thoughts and perceptions, while providing researchers with the opportunity to explore new ideas as they emerge during the interview~\cite{hiller2004,hove2005}. 

The interview protocol was centered around the research questions. Each research question inspired a specific set of questions, designed to trigger discussions on that topic. The protocol was collectively developed and revised several times by the authors of this paper. In the following, we provide a few examples of the questions in our interview protocol (which is fully available in our appendix~\cite{appendix_online}):

\begin{enumerate}[label=(\alph*)]
    \item RQ$_1$, Security in the SDLC: Can you explain to me how the software development process looks like in your current project? Can you tell me what your day-to-day work looks like? Where does security fit into your daily activities? 
    \item RQ$_2$, Security Practices: How do you or your team ensure application security in the current project? How do you prioritize the tasks, and what is the source of urgency?
    \item RQ$_3$, Security Tools: What security tools do you use and trust in your development pipeline? From a usability perspective, where do you see these tools need improvement, or in other words what do you miss in terms of tooling?
    \item RQ$_4$, Learning Resources: What type of security knowledge do you, or an engineer from your current team, need to perform in your current project? What do you do to ``learn security'' in your current project? How do you compare learning about security and learning other domain-specific technologies (e.g. learning relational database), in terms of difficulty and complexity?
    \item RQ$_5$, Challenges in applying security: What challenges do you currently face in applying security in your current project? How do you describe the difficulty of the security tasks you perform in your current project compared to the domain-specific tasks (e.g., to code functionality in the system/application)? What makes them challenging or difficult?
    
\end{enumerate}

We also followed two guidelines. First, whenever talking about security practices, challenges, and tools, we asked participants to focus on their current experiences, rather than on their general knowledge or beliefs about the topic. Moreover, we constantly asked them to illustrate their point of views with concrete examples from their current project. In practice, we incentivised participants to share stories. While we do not report the concrete stories in the paper to avoid any information leakage, they helped us to understand the participants' point of view. Finally, we also note that, due to the given nature of semi-structured interviews and variability in demographics and professional experience and expertise, not all the topics were discussed at the same length and level with the interviewees (e.g., for interviewees that are software developers, the interviews naturally focused more on the technical aspects of the software security; for interviewees that hold more managerial positions, the interviews naturally focused on the broader challenges that security brings to the software). We followed Glaser and Strauss’s recommendation and concluded the data collection when new data didn't add any value to the analysis, i.e., on reaching saturation~\cite{glaser1967the}.

The interviews were conducted via video conferencing (n=15) and in-person (n=1), recorded, and transcribed. Each interview lasted between 45 minutes to two hours, totaling 18.7 hours of audio recording. The first author conducted 12 interviews, and the second author conducted four interviews. These four interviews were conducted in Spanish and later translated to English.

\subsection{Data Analysis}

We performed open, axial, and selective coding in the transcriptions of the interviews, as commonly applied in empirical software engineering research~\cite{holsti1969content,thomas2011qualitative}. As a first step, the first author of this paper interpreted each discrete piece of information from the interviews, and assigned a code to it. Once the initial open coding was completed, the first author then established relationships between the emerged codes (i.e., related codes were grouped together, codes that could explain possible relationship or consequences were linked together). Finally, the first author then selectively assigned codes to the four main research questions of this paper.

The second author of this paper supported the entire analysis process, by revisiting and refining the code book together with the first author. The final observations that emerged out of the analysis were then revised and discussed with the third author of this paper. At any point of the process, disagreements were solved by means of revisiting the raw data until reaching a consensus.

\subsection{Reproducibility}

Our appendix~\cite{appendix_online} contains the code book with coded segments, assigned codes and interview protocol document. We do not release the raw interviews as they may contain sensitive information about the participating financial services companies and their projects.
\section{Results}
\label{label:result}

\begin{figure}
    \centering
    \includegraphics[width=1.3\columnwidth]{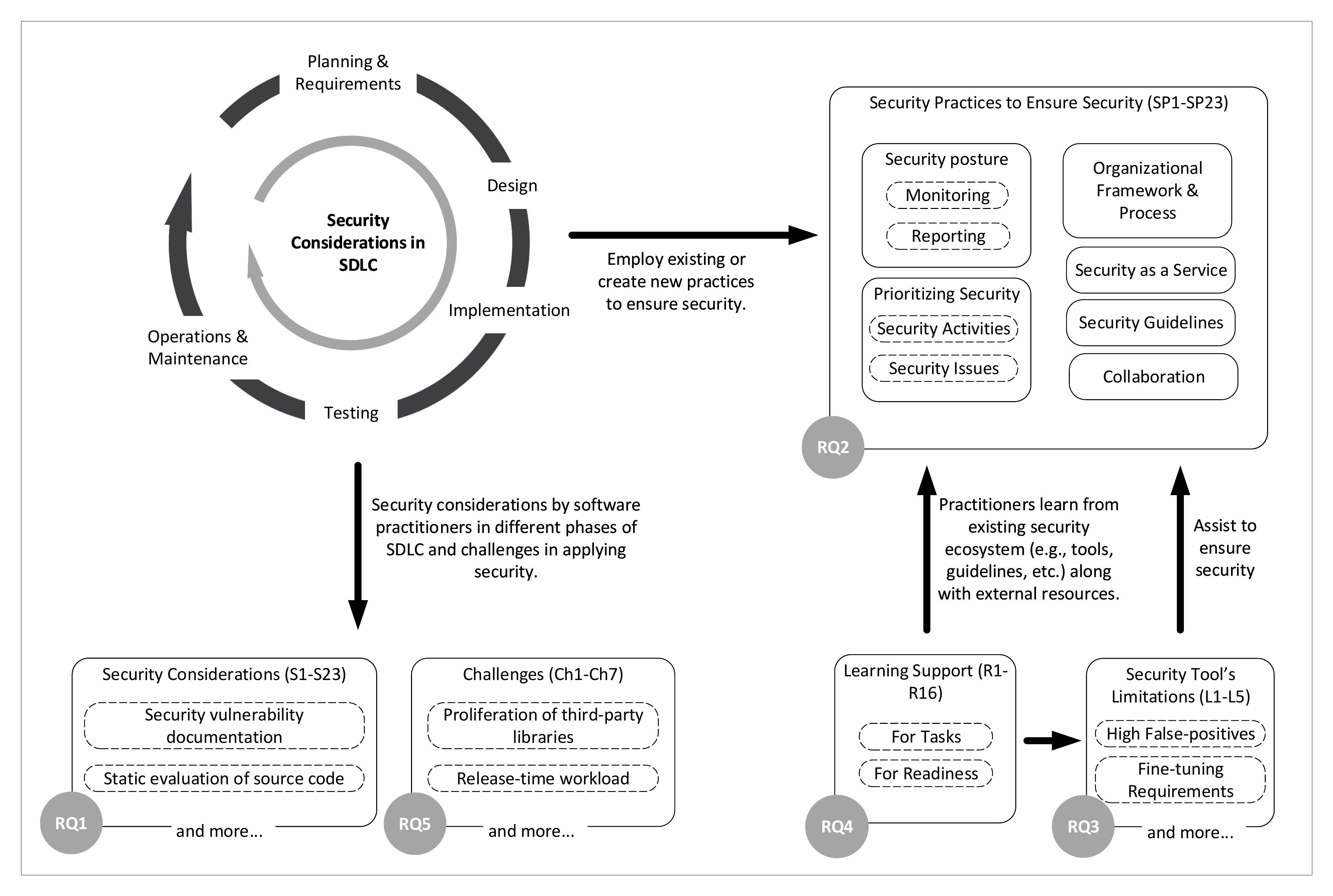}
    \caption{Overview of the results and examples of emerged categories according to the research questions (see Tables~\ref{tab:securityinsdlc} to~\ref{tab:learningandresources} for complete list). }
    \label{fig:resultsoverview}
\end{figure}

This section presents the main components of our findings from the interviews with software practitioners, which we illustrate in Figure~\ref{fig:resultsoverview}. Tables~\ref{tab:securityinsdlc} to \ref{tab:learningandresources} provide a complete list of emerged observations from the qualitative analysis. All interviewees discussed the security aspects from their current project and team's perspective; therefore, the reported results represent the security practices applied and observed at present. We discuss each of these categories in detail and the influential factors in the following sections.

\subsection{Security Considerations in the Software Development Life Cycle (RQ1)}
\label{lab:securityinsdlc}

We observed that the security considerations vary across development phases, which we discuss in each subsection. Sections~\ref{lab:planingAndrequirement} to \ref{lab:operations} present the findings according to the SDLC stages and Table~\ref{tab:securityinsdlc} lists them (S1-S23) for reference.
We highlight each SDLC stages' important security considerations and the interviews' key factual descriptions. 

\begin{table*}[htbp]
  \centering
  \scriptsize
  \caption{Incorporating security in the SDLC of financial services domain.}
\begin{tabular}{lp{25em}p{10em}}
\toprule
\textbf{ID} & \textbf{Security considerations} & \textbf{Participants} \\
\midrule
\multicolumn{3}{l}{\textbf{Planning \& Requirements}} \\
S1    & Secure from the beginning and secure by design. & P1, P2, P4, P5, P7-P13 \\
S2    & Products owners are more vigilant for security from the beginning. & P4, P7, P9, P10, P14, P15 \\
S3    & Security team is involved by-default in this phase. & P1, P13 \\
S4    & Security team can be involved on-demand by the development team. & P9, P10 \\
S5    & Identify security requirements and applicable security standards. & P2, P7, P11, P13, P15 \\
S6    & Prioritizing security efforts for the supported software artifacts from the beginning. & P2, P4 \\
\hdashline
\multicolumn{3}{l}{\textbf{Design}} \\
S7    & High volume of security related discussions. & P2-P4, P6, P7, P11, P12, P14, P16\\
S8    & Security team assist developers' implementation of security by choosing security technology/mechanisms. & P2-P4, P6, P7, P12, P14, P16 \\
S9    & Security architect plays the most important role in design phase. & P3, P4, P11, P12, P16 \\
S10   & Security team conducts threat modeling and security risk assessment activities in design phase too. & P2, P7 \\
\hdashline
\multicolumn{3}{l}{\textbf{Implementation}} \\
S11   & A dedicated team to ensure secure coding. & P2, P3, P7, P8, P10, P12, P13, P16 \\
S12   & The focus of developers is primarily on building feature. & P7, P9, P10, P11, P13, P14 \\
S13   & Static evaluation of source code (manual and/or automatic) is an important consideration in implementation phase. & P2, P5, P6, P8, P11, P13 \\
S14   & Multiple iterations of threat modeling and penetration testing during implementation phase. & P2, P16 \\
\hdashline
\multicolumn{3}{l}{\textbf{Testing}} \\
S15   & Security testing primarily depends on automated tools. & P1-P16 \\
S16   & The security issues are identified and prioritized in testing phase. & P1, P2, P4, P7, P10-P12, P14-P16 \\
S17   & The use of penetration testing is generally ad-hoc for critical features. & P1-P3, P11, P12, P14, P16 \\
\hdashline
\multicolumn{3}{l}{\textbf{Operations and Maintenance}} \\
S18   & Incident management process to handle the reported security vulnerabilities.  & P1-P16 \\
S19   & Organization should have a plan of action (playbook) to handle different types of security incidents. & P2, P16 \\
S20   & Multiple levels of escalation for security-related analysis. & P11, P15, P16 \\
S21   & Documenting security vulnerability is an important activity. & P3, P5-P7, P9-P11, P13 \\
S22   & Security reports are not easily comprehended by other teams. & P6, P7, P10, P11 \\
S23   & Application deployment in the production environment is equally important as security testing. & P1, P2 \\
\bottomrule
\end{tabular}%

\label{tab:securityinsdlc}

\end{table*}

\subsubsection{Planning \& Requirement}
\label{lab:planingAndrequirement} 
In software development, various stakeholders have manifold meetings to define the requirements and decide the product's priorities and goals during the planning and requirement gathering phases.

Financial services companies embrace the idea of ``\textbf{secure from the beginning and secure by design}'' \observation{} (P1, P2, P4, P5, P7-P13). The product is design from the beginning to be secure. It implies using security mechanisms and patterns as guiding principles for developers to included security from the beginning of their software development life cycles (SDLC). One of the interviewees (P4) summarized the general role of security teams starting from the requirement phase as: ``\textit{[...] the security [team] is used to the following: During the requirement phase, list the security requirements, and then later in production, test those requirements with methods like hacking}''. P11, an information security architect and product owner, emphasizes that considering security in the end is already a lost battle. He explains: ``\textit{[...] you might have lost the game in the beginning and now you're just basically trying where you have ten holes and you have two hands to cover those two holes and eight still always will be available for threat actors. You have to secure by design, and you have to secure from the beginning.}''.

We observe that the different stakeholders managing the product's overall development, such as the product owner and program manager, are already \textbf{more vigilant when it comes to security} \observation{} right from this phase (P4, P7, P9, P10, P14, P15). Although they are not security specialists, they should be aware of any risk acceptance, according to P7 and P10. Therefore, security and development teams help define the security requirements and later incorporate them into the product's architecture and design.

However, while companies think of security from the very beginning, the way security teams engage varies. More specifically, it can happen in two ways: by-default or on-demand. The \textbf{by-default} involvement happens when the requirement is embedded in the company's standard procedures and policies \observation{} (P1, P13). P1 exemplifies, ``\textit{Part of the standard (company's standard process) [...], is managing (including) a security advisor for each development. This security advisor needs to be involved in every development from the beginning.}'' On the other hand, security teams involved \textbf{on-demand} \observation{}, join the process once the development team identifies security-related requirements (P9, P10). For example, the development team first works with the product owner to find the customer requirement's feasibility during the planning and requirement phase. If at that point, a requirement is identified as security-related, the security team is engaged for further refinement.

Once a security-related requirement is identified,
the development team and the security team work together to \textbf{identify the known security standards} and \textbf{security patterns}\cite{Fernandez-Buglioni2013-dy} \observation{} that apply to the project (P2, P7, P11, P13, P15). For example, P2 pointed out that requirements are not just about application functionalities but also security considerations. 
Another participant, P13, pointed out that the security team helped his team to improve the permission checking code by introducing a known security standard. P13 says that, in a particular case, the suggestions offered by the security team not only helped them in having a safer permissions system in place but also ended up in easing the debugging process later on.

Another role of security teams is to support the development team in \textbf{prioritizing the security efforts of the different features and/or applications in the company's portfolio from the beginning} \observation{} (P2, P4). After all, different applications require different security efforts, and prioritizing helps companies optimize the allocation of security resources, e.g., automated tools, IT infrastructure, and human resources (P2). P4 exemplifies: ``\textit{[...] ``every once in a while'' or ``[...] only the functionality of the current sprint'' or ``the whole application'', this kind of decision needs to be made at the beginning. If not, you always do many and long tests, which take a long time and delay production and impact the business.}''

\subsubsection{Design}
\label{lab:design}
The design phase involves discussion with stakeholders on the design and architecture of the software artifact focusing more on the ``how'' rather than the ``what''.

We note that this phase is characterized by the \textbf{high volume of security-related discussions and questioning} \observation{} (P2-P4, P6, P7, P11, P12, P14, P16).
In our interviews, participants (P2-P4, P7, P10-P14, P16) pointed out many discussions about design-time security considerations. These include, decisions on security algorithms, encryption methods, selection of third-party libraries, usage of circuit breakers\footnote{https://martinfowler.com/bliki/CircuitBreaker.html}, selection of development frameworks, security testing, logging, approval of design and architecture, the security pipeline (integrated security checks in CI/CD pipeline) designing and implementing necessary security controls, making a threat model, ensureing secure data and code flow, and update or change the design or architecture.

The security team's role in the design phase is \textbf{to ease the developer's implementation of security primarily by helping them choose technologies and security mechanisms} \observation{} (P2-P4, P6, P7, P12, P14, P16). 
The development team takes necessary security-related inputs from the security team. There are differences of opinions, which are discussed positively in the meetings (P4, P7). For example, P14 illustrates a case where security experts gave him some suggestions on using Google Cloud in a much more secure way. P16 pointed out that based on the maturity of the project(s) in progress, he helps developers with design and technology choices. Participant P3 pointed out that the development team contacts him when they have to pass the security check. He exemplifies as ``\textit{They (developers) explain the architecture and the application. I collect all the diagrams the logic diagram and and also the infrastructure diagram and try to put those in a document. And also what security algorithms are they using if they're using TLS? Are they using transport level encryption methods? What kind are they using? I need to put that in the document. I need to also state if they're using third party software. I need to put all everyday level that you participate because they need to ensure that all the data that is transferred from one application to another application is safe.}''  

We observed that \textbf{the security architect plays an important role in the design phase} \observation{} (P3, P4, P11, P12, P16). The collaboration of development teams and product managers with the security architects is maximum in this phase. This is due to the fact that paying less attention to design and architecture can lead to major challenges later in the development life cycle. A security architect approves the architecture and design-time security requirements. 
P11 explained to us that architecture is one of the security pillars in his organization, and there is a dedicated team for architecture and design time security considerations. Moreover, the cost of security negligence can be very high in this phase, and it may lead to changes at the architectural level, as emphasized by P10 and P12. 
P10 exemplifies as ``\textit{[...], and then you have to deal with an architectural problem because of the data you are storing. And this kind of situation is a bit more difficult because you have to change architecture because of it.}''

Interestingly, two participants mentioned the \textbf{threat modeling and security risk assessment} \observation{} activities in the design phase (P2, P7). Threat modeling is a structured way of identifying and prioritizing potential threats to the system or software artifact and finding possible mitigation techniques.
Threat modeling and risk assessment in the design phase is about security preparations in advance. According to P2, with the help of threat modeling and risk assessment, their team was able to identify the security controls\footnote{https://owasp.org/www-community/controls/} to be implemented in advance. P7, a senior developer and lead, pointed out that 25\% of their design phase is dedicated to security, and they cannot afford security negligence for the financial services. He further exemplifies, ``\textit{When you do the risk analysis, you already know that this customer is very important, so we can't afford any negligence or damage here. So, the risk is high for us, which means that that dashboard becomes a CI rating 111 (high priority) because the integrity and availability are critical. We have to make sure that we have night support if things go down. We have to make sure that we have a lot of logging to know if it is a hack (security breach) or something else [...], So all these are thoughts that we have to put in our design phase [...] So this is like preparing for the security [...]}''.

\subsubsection{Implementation} 
\label{lab:Implementation}
During the implementation phase,
decisions related to security requirements from the previous requirements and design phase guide developers while writing and debugging their code. 

We note that the \textbf{large enterprises have a security team to ensure secure coding of the product under development} \observation{} (P2, P3, P7, P8, P10, P12, P13, P16). Such a team (the ``secure coding team'') is responsible for security tasks such as manual code review, monitoring third-party dependencies, handling tool alerts, and monitoring the implementation phase progress (P2, P5, P6, P8, P11). In the case of P5, he is the only security analyst to support the developers from a security perspective in the entire organization. 

While \textbf{the focus of software developers is on building features} \observation{}, developers are still aware of security, but that security is not their primary concern (P7, P9, P10, P11, P13, P14). For example, P9 pointed out that he never got a separate security task, but that could be a part of a development task, such as ``creation of endpoint or a controller'' or ``removal of personally identifiable information from logs''. Another participant, P11, explains that the development team understands that they have to implement the backlog features in a secure manner. 

The \textbf{static evaluation of source code} \observation{} is an important security consideration in the implementation phase (P2, P5, P6, P8, P11, P13). This is done both manually and with the help of automated tools. The security tools can be embedded in the integrated development environment (IDE) to help developers find bugs while writing code. P2 says: ``\textit{Another important task is the static evaluation of source code. This would be done during the development, preferably. So while developing, the programmer can identify these vulnerabilities. It is a manual process, but nowadays, there are many tools to automatize this, which are integrated into the IDE. The IDE is the software that is used to develop. So, while programming, the IDE's integrated functionality automatically detects faults}''. The security team on the other hand has continuous work in this phase because it involves code review on the requests from developers or alerts from security pipeline. P8 exemplifies as ``\textit{our main responsibility for now in is in the continuous integration area, so it's actually more into the code reviews. Usage of 3rd party libraries and open source components in our software. So yeah, this is actually our main responsibility as a SECO (secure coding) team}''.

Finally, two of the participants pointed out that they conduct \textbf{multiple iterations of threat modeling and penetration testing} \observation{} in this phase (P2, P16). The output of threat modeling scopes the security issues to be covered during pen-testing. P16 pointed out that they conduct threat modeling and pen testing even if some features are not fully developed, especially to find the possible attack vectors.

\subsubsection{Testing}
\label{lab:testing}

In the testing phase, the development team focuses on finding out if the code works according to the customer's requirements. In general, the development team builds a test plan with decisions such as allocating the necessary resources, defining the test cases, how to conduct the testing, etc. In the context of security testing, it consists of static and dynamic testing. Both static and dynamic testing can have a manual and automated approach.

The participants pointed out several security considerations in the testing phase, including vulnerability scanning, secure code review, threat modeling, penetration testing, application verification testing, ethical hacking, and using static and dynamic code analysis tools.

We observed that all the participants emphasized \textbf{the use of security testing tools} \observation{} in this phase (P1-P16). As discussed for the implementation phase (Section ~\ref{lab:Implementation}), the security team can embed these tools in the continuous deployment pipeline and the developers' IDE. While the IDE integration helps during the implementation phase, the code scanners and automated test tools in the security pipeline help in the testing phase with automated scanning and testing. 

\textbf{The security issues are identified and prioritized} \observation{} in the testing phase (P1, P2, P4, P7, P10-P12, P14-P16).
The security issues are identified via security tools, manual code review, threat modeling, penetration testing, etc. 
In practice, this security testing helps the security analyst to target risky parts of code for code review. The code review process, together with OWASP's top 10, helps prioritize the identified security issues. Once the code review report is ready, the security analyst collaborates with the product managers to develop the security controls and mitigation techniques. P16 exemplifies the process as, ``\textit{We start from threat modeling, usually to try to understand who the actors could be. [...] We look at the exposed endpoints. [...] Then we try to understand what is the biggest risk and what are the critical components? Then we tried to map out attack vectors, how malicious users could inject all kinds of stuff that would go through the application. Then we cherry-pick specific areas in the code because you can't cover everything [...] Then, I categorize it by OWASP top 10, and give practical suggestions to mitigation [...]}''.

\textbf{The use of penetration testing is generally ad-hoc for critical features} \observation{}. Seven participants mention penetration testing during the interviews (P1-P3, P11, P12, P14, P16). The outcome of the code review in threat modeling helps to scope the application for penetration testing. Penetration testing involves more manual effort, and it is usually the last manual check before release. Participant P11, an information security architect, defines penetration testing as one of the security products available for the development teams. 
P11 explains: ``\textit{Because penetration tests are requested ad-hoc in a queue based mechanism where on a certain month, there are 10 critical applications in the organization which want their publicly exposed changes or features on their websites tested. It could be a month where only one of such applications is ready to release a major security-relevant feature.}'' .

\subsubsection{Operations \& Maintenance}
\label{lab:operations}
 
In the context of security the operations and maintenance phase is mainly about security incident management and application deployment.

In our interviews, we observed the financial services companies having an \textbf{incident management process to handle the reported security incidents} \observation{} (P1-P16). The incident management process has a step-by-step approach to handle security incidents. 
Once the incident is identified, the priority is to contain the incident (P1, P15). 
We note a high use of different tools such as Confluence, ServiceNow, Power BI dashboard, and JIRA, supporting the tracking of the incident.

Not all security incidents are severe to require a high level of attention. In other cases, though, such incidents are big enough to need a dedicated person to monitor the incident's handling from start to end. Such a high-priority incident or cyber-attack is taken as a crisis by the financial services companies. According to P2 and P16,  \textbf{organization should have a plan of action (playbook) to handle security incidents} \observation{}. 
Moreover, according to P2, the plan should be granulated by the type of attack of the incident (e.g., ransomware, denial of service). P2 exemplifies: ``\textit{First, what you do is identify the type of incident. Then you start using these playbooks [a plan of action], depending on the attack. For example, a playbook for a Denial of Service attack is different from one for ransomware and is different for a non-authorized intrusion, which caused a loss of data. For each of these, it is defined how to act}''.

The financial services companies also have \textbf{multiple levels of assignment and escalation} \observation{} for handling security incidents, such as incident monitoring, data extraction, threat modeling, forensics, and recovery experts (P11, P15, P16). P11 mentions that for critical issues, they have a concept of the emergency lane for quick response: ``\textit{If a new vulnerability comes out on the web or the Internet, for example, there was recently Citrix and DNS related very high severity vulnerabilities for which patches or mitigation was also released. You need to pick them in what we call emergency lanes.}". 


The \textbf{documentation of the vulnerabilities
}\observation{} that are observed in the system is also an important event during security incident management (P3, P5-P7, P9-P11, P13). We observe different responses on the comprehension of security report documentation. P5 mentioned that there is no formal way of documenting security vulnerabilities and P13 emphasizes the need for auditable documentation for backtracking. According to the participants P6, P7, P10, and P11, \textbf{security reports are not easily comprehended by other teams} \observation{}, such as developers. For example, P7 reported them as often too big to comprehend, and P6 points out that sometimes developers cannot comprehend why a bug is tagged as a security bug. P6 exemplifies as, ``\textit{In many cases it's not quite clear to the developers why this is wrong. [...] Why is this Patch from a security point of view? (tagged as security bug)}''.


Another aspect of reporting is the post-mortem of security incidents, where the technical analysis of the incident is communicated along with the recommendation for improving the overall security posture of the organization. P2 mentioned a real-life example of handling ransomware. Such a security incident requires many days of teamwork at multiple levels (technical, tactical, and strategic). It may bring big organizational changes and attitudes towards software security. P2 says: \textit{``In the end, security incidents leave you with many lessons learned both for the organization and individuals''}.

Finally, two participants (P1, P2) mentioned that \textbf{application deployment in the production environment is equally important as security testing} \observation{}. The application deployment defines the environment constraints and security measures to protect the application against potential attacks and monitor the application. 
The infrastructure (e.g., firewall, VPN, application proxy) of the company also plays a role in protecting the application. For example, P2 explains: ``\textit{[...] Same for the deployment, there is a strict process that needs to be followed before deploying any application. Once we are in production, we have services to monitor the application. For example, a well-known product is Akamai. It provides different security stages; one of those is the anti-DDOS, which prevents distributed denial-of-service attacks on the internet. It also provides the services of GWAVA, which is a web application firewall. On different levels of the application, we place rules to be able to recognize possible attacks.}''

\subsection{Ensuring Security (RQ2)}
\label{lab:ensuringsecurity}

This section discusses the real-life security practices that software practitioners and their teams adopted to ensure security. We identified a total of 23 security practices (numbered as SP1 to SP23) that we group into six higher-level categories as shown in the Table~\ref{tab:securitypractices}. These categories emerged as a consequence of grouping similar security practices. For example, the category ``Security as a Service'' contains the security practices, which are the security services offered by the security team to developers.

The six categories are: 1) The reporting \& security posture (SP1) category groups the participants' responses that discuss the security posture or indicate the reporting to monitor the security posture; 2) The organizational framework and process category (SP2 to SP6) groups the participants' responses representing a framework or a process within the organization that everybody adheres to ensure security; 3) The security guidelines category (SP7 to SP10)  groups the practices participants use to conduct their daily-routine securely; 4) The security as a service category (SP11 to SP20) groups the security practices which are offered as a service or product by the security team (or expert); 5) The prioritizing security category (SP21) lists one practice, and involves prioritizing security-related activities in their current project; finally 6) the collaboration category (SP22 to SP23) groups the practices related to the collaboration among software practitioners.

In the following we present each category with identified security practices in detail and highlight the participants' key factual descriptions.

\begin{table*}[htbp]
  \centering
  \scriptsize
\caption{The security practices used in the financial services domain.}
\begin{tabular}{lp{25em}p{10em}}
\toprule
\textbf{ID} & \textbf{Security Practice} & \textbf{Participants} \\
\midrule
\multicolumn{3}{l}{\textbf{Reporting \& security posture}}\\
SP1  & Monitor and act to ensure a good security posture.&  P6, P7, P9, P11-P16 \\
\hdashline
\multicolumn{3}{l}{\textbf{Organizational framework and process}}\\
SP2  & Create a framework for organizational security needs. & P3, P4, P12, P14-P16 \\
SP3   & Have an organization level security awareness program. &  P4, P8, P10, P12-P16\\
SP4   & Coach developers to become security champions within the development team. & P7, P11, P16 \\
SP5   & Include security in sprint planning. &  P6,P16 \\
SP6  & Create an incident management process. &  P1-P16
\\
\hdashline
\multicolumn{3}{l}{\textbf{Security guidelines}}\\
SP7   & Identify the security requirements and applicable industry best practices. & P3, P7, P9, P10, P12-P14 \\
SP8   & Leverage the \textit{OWASP (top 10}\footnote{https://owasp.org/www-project-top-ten/}, \textit{OWASP ASVS}\footnote{https://owasp.org/www-project-application-security-verification-standard/}, Cornucopia\footnote{https://owasp.org/www-project-cornucopia/}) resources. &  P2, P4, P5, P8, P10, P12, P15, P16\\
SP9   & Follow the third-party's framework-specific security guidelines. &  P9, P13\\

SP10   & Define clear internal security guidelines. & P3, P4, P9, P13, P14, P15\\
\hdashline
\multicolumn{3}{l}{\textbf{Security as a service}}\\
SP11  & Create a security pipeline to incorporate security tools, checkpoints, and controls. &  P6-P10 \\
SP12  & Perform static and dynamic analysis on code by leveraging automated tools. &  P1-P16\\
SP13  & Manage and monitor security vulnerabilities for the supported software artifacts.  & P6-P8, P11, P13 \\
SP14  & Manage and monitor third-party libraries for security vulnerabilities.  & P4, P5, P8, P14 \\
SP15  & Perform secure code review to identify security flaws in code.  & P2-P8, P10-P13, P16 \\
SP16  & Perform regular risk assessments to identify risk factors. &  P7, P11, P12, P14, P15\\
SP17  & Perform threat modeling to identify the attack vectors and weak areas in code.  & P2-P5, P7, P10, P11, P14-P16 \\
SP18  & Perform pen-testing for deeper inspection. & P1-P3, P11, P12, P14, P16 \\
SP19  & Perform forensics on traces left after a security incident.&  P4\\
SP20  & Monitor the security indicators and identify the indicator of compromise. & P1, P4 \\
\hdashline
\multicolumn{3}{l}{\textbf{Prioritizing security}}\\
SP21  & Prioritize the security-related activities at the organization and project level. & P3, P5-P8, P11-P14\\
\hdashline
\multicolumn{3}{l}{\textbf{Collaboration}} \\
SP22  & Security team should maintain a good rapport with other teams. &  P1, P4, P6, P13, P15, P16\\
SP23  & Teamwork and team dynamics are important for security. &  P2, P4, P5, P7, P8, P11, P16\\
\bottomrule

\end{tabular}%
\label{tab:securitypractices}

\end{table*}

\subsubsection{Reporting and Security Posture}
\label{lab:securityposture}

The security team \textbf{monitor and act to ensure a good security posture} \practice{} (P6, P7, P9, P11-P16). The security posture is an indicator of the health of the organization's overall security. Reporting helps monitor software development projects' compliance and the organization's current state of security or security posture. P11 explains: ``\textit{We provide oversight on the security posture by identifying security issues or vulnerabilities in the organization's IT landscape through various products [...]}''. P11 further mentions that their security teams support more than 3000 applications. Based on the team's reporting requirements, there are different reporting levels (such as detailed, aggregated, and overview) and reporting channels for each team. The regulations or compliance for the financial service companies are high (P6, P7, P12-P16). P16 explains: ``\textit{in a bank there is no shortage of compliance. I would say in a bank; there is more compliance than actual coding going on.}'' P16 also points out that General Data Protection Regulation (GDPR\footnote{https://eur-lex.europa.eu/eli/reg/2016/679/oj}) has also diverted the attention of managers and executives towards the security posture of the applications. 
During our discussion, the participants pointed out that the security posture of the organization depends on many factors, such as the number of security issues, security issues per cluster, incident response time, the severity of the security issue, affected servers, type of vulnerabilities, affected applications, top security defaulters, cause of vulnerabilities, vulnerability trends, etc.



\subsubsection{Organizational Framework and Process}
\label{lab:organizationalsecurityframework}
We observed that the real-life day-to-day software development practices in the financial services technology are conducted in an agile manner. According to P2, agile methodologies allow easy integration by delegating security to security teams. On the other hand, one of the difficulties we observe is that the agile methods can be too fast to incorporate security fully (challenge Ch3, discussed in Section~\ref{label:competingwithagilemethod}). 

However, while agile methodologies can easily integrate security concerns, a strong security framework is still required. In our interviews, many participants point out their existing \textbf{organizational framework} \practice{} for security needs (P3, P4, P12, P14-P16). P12 pointed out that the first thing they do to ensure security is to have a coherent architectural framework for security. The framework incorporates the priorities, goals, risks, internal security guidelines, compliance requirements, and other organizational security requirements (P3, P4, P14). P16 describes his organization
as agile bottom-up and has a well-defined security framework: ``\textit{[company] is very unique [...] we're structured like Spotify\footnote{https://www.atlassian.com/agile/agile-at-scale/spotify} so we're the first and maybe the only bank in the world that is agile from the bottom up. We have independent squads and then tribes. A tribe is a kind of division. [...] each tribe should be self-sufficient and they should have their own security architects and security operations and security pipelines.}''.


Participants (P4, P8, P10, P12-P16) pointed out that they are part of an \textbf{organizational level security awareness program} \practice{} either as a participant or contributor. Such programs offer periodic events where security vulnerabilities and good security practices are discussed. Organizations have a budget allocated for such activities (P8). P14, a product owner and a data scientist, pointed out that constantly communicating security topics helps to raise security awareness within the organization. 
P16 pointed out that training programmers is a must for security, and he is currently driving this initiative in his organization. He emphasized that he makes the training activities a fun event for developers: ``\textit{If I don't stir the wrenches, they won't cooperate, and they'll just think I'm another compliance nuisance and treat me as such. So what I do is, aside from training them, I do fun activities. At the end of the training, they must hack a casino and win points [...]}"

Participants (P7, P11, P16) mentioned the concept of a security champion, where one or more senior programmers are trained to become the security champion for the team. A \textbf{security champion within the development team} \practice{} acts as a bridge between the security team and developers. Some of the security champion's responsibilities include being first point of contact for security issues for developers, conducting secure code reviews, applying defensive programming techniques, and being the liaison of the team with the security architect (or security team). The concept of security champion supports the security considerations S7 (Section~\ref{lab:design}) and S12 (Section~\ref{lab:Implementation}) because developers can focus on feature development and security team gets a direct point of contact within development team.  P16, in particular, currently provides the necessary training for developers to become security champions. He further pointed out that the basic application security training for the security champion role is just for a few days, but it is a continuous learning process to establish a common language with the security team.

Two participants (P6, P16) pointed out that they always have \textbf{some space for security in the sprint planning} \practice{}. According to P6, this is a good way to integrate security in the process. A post-sprint security audit is another way to include security (P16). Integrating security in the development process can help to support security considerations and challenges in SDLC, such as, S6 (Section~\ref{lab:planingAndrequirement}), S22, S23 (Section~\ref{lab:operations}), and Ch6 (Section~\ref{lab:challenges}) etc. According to P6, this could be an investment in the team members to learn security. He explains: ``\textit{[...] within our Sprint planning and continuous workload, we try to always integrate a part which just purely has to do with security with the with improvements and research. [...] it can be just an investment in the team members [...]}''.

Finally, participants (P1-P16) mentioned the \textbf{incident management process} \practice{} (discussed in S18, Section~\ref{lab:operations}), which involves assigning the priority to the reported incidents, responding and solving based on the assigned priority, and allocating the necessary technical and management resources. The incident management process's goal is to avoid or limit the possible damage due to a security breach. Additionally, the lessons learned while handling the security incidents help the organization prepare better for future incidents.

\subsubsection{Security Guidelines}
\label{lab:securityguidelines}
12 participants pointed out various security guidelines they follow to ensure security.

Software practitioners put efforts into \textbf{identifying the security requirements} \practice{} (discussed in S5, Section~\ref{lab:planingAndrequirement}) that apply to their projects (P3, P7, P9, P10, P12-P14). Once the requirements are identified, security considerations can be discussed and adopted to help in other phases of software development. 

In our interviews, \textbf{the OWASP resources and guidelines were leverage by almost all the participants} \practice{} (P2, P4, P5, P8, P10, P12, P15, P16). More specifically, the OWASP top 10\footnote{https://owasp.org/www-project-top-ten/}, OWASP ASVS\footnote{https://owasp.org/www-project-application-security-verification-standard/}, Cornucopia\footnote{https://owasp.org/www-project-cornucopia/} as well as other resources related to application security guidelines, and secure coding practices. According to the participants, such resources help them in protecting themselves against popular security vulnerabilities. 

Two participants (P9, P13) pointed out the \textbf{third-party framework-specific guidelines} \practice{} in case the application makes use of a framework with published security practices. For example, P9 pointed out that their application uses the Spring framework, and their internal security guidelines recommended to adhere to Spring security guidelines.\footnote{https://docs.spring.io/spring-security/site/docs/5.0.x/reference/html5/}

Along the same lines, many participants (P3, P4, P9, P13, P14, P15) point that \textbf{well-defined internal security guidelines} \practice{} help them to ensure security. Participants provided many aspects where security guidelines are followed, such as handling application logs for personal information, handling customer and company data, data collection from internal or external teams, or documentation during a collaboration (e.g., security checks for projects in different phases of development). The security team expertise is utilized by developers in different phases of SDLC (e.g. S8, Section~\ref{lab:design}) and an ambiguity can invite conflicting scenarios. For example, P4 shared a scenario where the product owner reaches out to the security team to white-list some security vulnerabilities so that product release is not blocked and promises the security team that vulnerability will be immediately looked into post-release. If a security vulnerability is exploited, the business will come back to the security team asking why the identified issue was not fixed. This is one reason organizations need to define clear internal security guidelines, and everyone should adhere to them. The security guidelines ensure that at least critical security vulnerabilities are detected before reaching production (P4).  

\subsubsection{Security as a Service}
\label{lab:securityasaservice}
In our interviews, we observed that \textbf{security teams operate as consultants} (P6-P8, P10, P11, P13). For example, the participants P7, P10, and P13 point out that they invite the security teams during the different development activities (e.g., software design and architecture level discussions) for their inputs on security-related issues or explore the security aspects. Similarly, P6 and P11 point out that their role is less related to development but more on the consulting side. P11, an information security architect, says, ``\textit{We don't manage applications; we are independent assurers.}". He further explains that his team supports thousands of applications in the organization, but they are not involved in developing or managing the applications. We can infer from the data that security teams, in general, offer security as a service to other teams. Overall we observed in our discussions that the security team is approached for security issue's advice, e.g., ``is this a good way to do it?'', approval for tools and code, permission issues, data exposure issues, etc. 
We categorize those security practices as ``security as a service''.


The software practitioners use the \textbf{security pipeline} \practice{} for various security related activities (e.g. implementing security checkpoints) (P6-P10). 
A security pipeline is a well-designed CI/CD workflow to include security practices and tools. The automation and integration of development-related activity within the standard pipeline allows the security team to support and manage multiple projects in different development phases. For example, security testing primarily depends on automated tools (S15, Section~\ref{lab:testing}). P8 explains: ``\textit{our responsibility is to make sure that our applications are actually not vulnerable to such a known vulnerability. This is where we make sure that we, for example, build up certain custom rules or customization in the standard pipeline (CI/CD workflow) and the software development process to detect this kind of issue as early as possible [...]}''.

The security infrastructure to support applications (sometimes in large numbers) requires both manual and automated efforts. We observe a high dependency on automated tools (S13, S15 discussed in Sections~\ref{lab:Implementation} and~\ref{lab:testing}). 
The participants highlighted more use of \textbf{static analysis tools} (P1-P16) compared to \textbf{dynamic analysis tools} (P1-P4, P6, P7, P11, P12, P14) \practice{}. Many important aspects of security assurance, such as bad security practices, compliance, and validating general security practices, depend on the organization's static analysis capabilities. For example, P1 explains: ``\textit{Under static (analysis) would fall revising the source code with an automatic tool, to check for general security practice, but also checking for compliance with the banks' requirements..}''
We observed that nine participants mention the use of dynamic application security testing (DAST) tools to ensure security. 
According to P2, dynamic analysis tools are not fully automated and need manual intervention to validate the results. 

Many participants (P6-P8, P11, P13) pointed out the role of the security team in \textbf{vulnerability management} \practice{} of supported software artifacts. This involves the identification and classification of security vulnerabilities (manual and automated), continuous monitoring, and improving custom rules within security tools to detect vulnerabilities.
Software developers and the security teams, monitor and handle security alerts generated from the security pipeline.
Participants pointed out security tools such as vulnerability scanners and vulnerability reporting tools, are part of the vulnerability management process. They also customize and configure the tools according to project requirements (P6-P8). According to P6, vulnerability management is part of continuous maintenance (Section~\ref{lab:operations}). When there is some new vulnerability to handle, his team create custom rules for the security tools. These custom rules are also monitored for false-positives (P7). If some rule gives too many false positives, developers lower the severity or get approval to remove the rule (P7).

The \textbf{management and monitoring of third-party libraries for security vulnerabilities} \practice{} is also one of the major activities of the security team (P4, P5, P8, P14). According to P4:
``\textit{SCA (software composition analysis) tools are a great way to accompany this. [...] Your own software could be very secure, but if you call other libraries, such as a jquery, and it is an older version, or there is some other vulnerability in that library, that means that your software is now also vulnerable}''.

The security team performs \textbf{secure code reviews} \practice{} (manually and automated) (P2-P8, P10-P13, P16) to identify security flaws in the code.
P13, a developer, explains: ``\textit{You don't select your reviewers. If it’s security-related, at least one of the reviewers will be somebody from the security team}''. Static evaluation of source code is an important security consideration (S11, S13 in Section~\ref{lab:Implementation}) in financial services companies. 
Participants mentioned that manual secure code review is conducted on the pull-requests. The pull request is accepted or rejected, or additional comments are provided by the security expert (P5, P7, P12). The static analysis tools help to analyze the code automatically and generate alerts. Security team also reviews the post-automation alerts for existing or new security vulnerabilities (P8).  

The security team performs \textbf{risk assessments} \practice{} (P7, P11, P12, P14, P15) of the supported software artifacts. 
Risk assessment considers the threat model (S14 in Section~\ref{lab:Implementation}) and application/platform-specific factors (e.g., ranking or priority) (S6 in Section~\ref{lab:planingAndrequirement}) to identify the security controls for mitigation. P11 explains: ``\textit{some are on the assessment side of the risk, based on some industry-standard frameworks like IRAM 2\footnote{https://www.securityforum.org/tool/information-risk-assessment-methodology-iram2/} where you take into account the threat model or profile of a certain application or certain platform and then based on that you arrive at the kinds of controls that should be in place and how effective those security controls are [...]}''. 

Some of the security activities performed by security team such as, \textbf{threat modeling} (S10, S14 and S16, in Section~\ref{lab:design}, \ref{lab:Implementation} and \ref{lab:testing}) \practice{}, risk assessment and \textbf{penetration testing} (S17, Section~\ref{lab:testing}) \practice{} depend on each other. The output of threat modeling help assess the associated security risks, and secure code review during the threat modeling provides the potential code sites to target during penetration testing (P16). 



One of the participants pointed out the \textbf{forensics analysis} \practice{} by the security team for deeper investigations (S20 in Section~\ref{lab:operations}) on a security incident (P4). The forensic analysis involves detailed investigation to find the root-cause and consequences of a security incident. 
P4 explains: ``\textit{We have forensic analysts, who are the ones that face the identification of the traces the attacker left within the system. [...] Then we have another team that takes care of the recovery. So let’s say they have attacked me, and my systems are out of service, how can I recover to return to the operation? It is a mix of different profiles [...]}''.

The security activities are increased or decreased according to security indicators (P1, P4, P16). The security team \textbf{monitors the security indicators} \practice{} for the supported applications to assess the overall security and progress. P4 explains: ``\textit{There are some indicators that you want to monitor to be able to keep track of the security progress. At first, these are very basic, e.g., how many applications are connected to my security tool, how many vulnerabilities do we detect every sprint, what is the level of [...] that we need for every application? Now you have these indicators, and as soon as you see them rising, it is time to increase security.}''

\subsubsection{Prioritizing Security}
\label{lab:prioritizing-security-tasks}
In our interviews, we observed that \textbf{security-related priorities} \practice{} are discussed at two levels. First, prioritizing security-related activities (S1 and S6 in Section~\ref{lab:planingAndrequirement}), such as identifying security requirements, secure code review, and security testing. Second, the prioritization of the security issues (S18 in Section~\ref{lab:operations}).

The teams' decisions on security-related activities are taken collectively during meetings (P5, P7, P9, P10, P12, P13). We note that a few stakeholders have more power to decide the teams' priorities, such as the technical lead, the product owner, or the security head (P5, P7, P9, P13). In other words, they can decide on undertaking a security activity (e.g., threat modeling) or completely skip it based on the circumstances (e.g., workload, release date, etc.).  P7 explains: ``\textit{[...] the priorities are set by the team. So in the refinement, we have a clear discussion about what is 1 what is 2. And it's not only security (team) that decides. It's basically normally the business value which decides, and security is considered as business value for us}''. 

The prioritization of security issues depends on the technical and organizational factors:

\textbf{Technical factors:} The priority at the technical level is either assigned automatically by the security tools or manually by evaluating the type, exploitability, and risk associated with the security issues (P8, P10, P11, P13). The security tools dispatch the found security vulnerabilities and assign the priority based on a pre-defined configuration. P8 explains: ``\textit{[...] depending on the criticality and the risk, for instance, we have two issues or two vulnerabilities that we would like to address across the whole development [...] a vulnerability, if it got exploited in the wild, then it might just have very minimal information disclosure. However, on another issue, if it would get exploited in the wild, it will reveal private information. So based on the assessment of the risks and also the type of vulnerabilities, we put the priority on there}''. 
    
\textbf{Organizational factors:} The factors such as release date of the project, application ranking, issue assigner, and compliance are measured at the organizational level. These factors also contribute to deciding how the security issues will be handled (P3, P6, P7, P11). A project near the release date has more priority for the stakeholders, such as the product owner, program manager, etc. P3 mentions that he needs to give quick attention to vulnerabilities during release time. 

Similarly, some applications or parts of the applications may be more critical for security. In such cases, the application's issues are given more priority over others. For example, the pen testing team will prioritize the tasks related to the application's front-end components rather than back-end components (P11).


Finally, prioritizing security issues also depend on the assigner of the security issues. The developers are the main customer for the security team; therefore, developers' tasks are prioritized first (P6). The escalation potential of the security issues also provides a priority boost. P13 pointed out that some tasks can become a high priority if not attended in a short time; hence they need to be prioritized over others.

\subsubsection{Collaboration} 
\label{lab:collaboration}
Developers are the main customers for the security team. Therefore, ensuring security requires effective collaboration among different roles and teams (P1, P3, P4, P6, P10-P13, P16)).

\textbf{Communication skills are fundamental for security teams}, leading to better collaboration and knowledge transfer with the development team \practice{} (P1, P4, P6, P13, P15, P16). 
Moreover, the social skills build a rapport with developers and help to share knowledge effectively. P4, a cyber-security manager explains: ``\textit{As soon as the security team abuses this power and uploads thousands of requirements, the development team will see the security team as the police that doesn’t help them but causes them to fall behind on everything.}''
Carefully explaining the consequences of specific security breaches is important to make developers more confident about the solution (P16).



\textbf{Teamwork and team dynamics are important for security} \practice{} (P2, P4, P5, P7, P8, P11, P16). Large enterprises have many security teams (secure coding team, penetration testing team, etc.), and they need to collaborate to ensure security. Security is not just a technology or tool; people and processes are part of it as well (P2). Security needs a high amount of manual collaboration, and there are challenges with it. P7 explains: ``\textit{At the design time, there is a lot of communication [...] there is always a challenge because [...] there can be differences of opinion.}'' P11 pointed out that the pen testing team needs threat-based scoping as input to be effective, provided by the team responsible for threat modeling of the application in question. P8 expands on a collaboration challenge within the security teams. He pointed out that each security team is doing their tests, but complete visibility (on findings or assessment) from all the teams is lacking.



\subsection{Security Tools (RQ3)}
\label{lab:securitytools}

In our interviews,
participants mentioned a plethora of different tools with numerous capabilities, e.g., static and dynamic analysis, Active Directory-based authentication, single sign-on, patch-aware scanning, information verification, scanning of secrets, docker image scanning, software composition analysis capabilities. In this section, we discuss the 
the current limitations of security tools in the context of financial services. Table~\ref{tab:securitytools} shows a summary of our findings.


\begin{table*}[htbp]
  \centering
  \scriptsize
\caption{Security tool's benefits, selection and limitations}
\begin{tabular}{lp{25em}p{10em}}
\toprule
\textbf{ID} & \textbf{Observation} & \textbf{Participants} \\
\midrule
\multicolumn{3}{l}{\textbf{Limitations}}\\
L1   & High false-positive rate & P2-P5,P7,P8,P11,P16 \\
L2   & Fine tuning requirements &  P1, P4, P7, P8, P16\\
L3   & Integration capabilities &  P1,P3,P4,P6,P11,P15\\
L4   & Lack of root-cause and intuitiveness capabilities & P4, P6,P8, P16\\
L5  & Non-functional requirements in security tools &  P3, P5, P6, P8,P9, P16 \\
\bottomrule

\end{tabular}%
\label{tab:securitytools}

\end{table*}

\subsubsection{Security tool's limitations}
\label{lab:securitytoolslimitations}
In the following, we present the key factual description by the participants on the security tools' limitations.

\cooltitle{High number of false-positives \limitation{}:} Several participants (P2-P5, P7, P8, P11, P16) point out that false positives are one of their great concerns when using automated security tools. 
This is one of the pain areas, especially for security engineers and developers, as a lot of time is invested in handling false positives (P7). P7 pointed out that developers at his company often spend around 5\% of their time dealing with false-positives, which collectively translates to a large number of person-hours a month. We also observed that the security experts are more frustrated with the high false-positive rates. P4 mentions that when you first run the tool out of, e.g., 200 alarms, only 40 may be legitimate. Moreover, a big code base amplifies the number of alerts by many folds.
P16 explains: ``\textit{You can have static code analysis running on the code while you're still coding it in your IDE, such as Visual Studio or IntelliJ. And then you can get suggestions, but they are highly inaccurate. We found out without mentioning vendor names that they reached 96.6\% false positives.}''


\cooltitle{Fine-tuning the tools \limitation{}:} The security tools require a good amount of effort for their initial fine-tuning (P1, P4, P7, P8, P16). P8 pointed out that fine-tuning is required for almost all the static analysis tools available in the market. In a large enterprise, the threats are monitored daily, and fine-tuning of security tools is a continuous process (P1). According to P16, no organization has fine-tuned the security tools because code changes are quick, and fine-tuning these tools becomes impossible. P16 explains: ``\textit{[...] no organization is able to fine-tune these tools because if you have over 1000 programmers, it's not scalable because you have 10s of projects and each project needs its own fine-grained policy on the tool that is used centrally, and as they change the code and frameworks being used, they need to keep updating the policy of the tooling, and so it's set to fail.}''

\cooltitle{Integration capabilities \limitation{}:} 
Tools may have good technical ability, but they may not follow the industry standards. Hence, the integration capability is a concern (P1, P3, P4, P6, P11, P15). Security tools can help detect security vulnerabilities; however, the developer(s) of these tools may overlook the integration needs. The integration requirement can be in the context of business tools (e.g., Confluence\footnote{https://www.atlassian.com/software/confluence}), development pipeline, development environment, etc., as pointed by P4 and P1. P11 explains: ``\textit{[...] certain tool spits out vulnerabilities and marks them based on their industry knowledge in a certain category. Whereas that categorization is not in line with your organizational structure, organizational contracts, and it also doesn't allow you to customize it.}".

\cooltitle{The root-cause and the intuitiveness of the tools \limitation{}:} Participants P4, P6, P8, and P16 point out the need for more accurate and intuitive (user friendly) security tools. P16 pointed out that software composition analysis tools may tell you if there is a vulnerability; however, they are incapable of further digging to help the practitioners know where the vulnerable code is being used in the application. Similarly, P8 pointed out that tools should be intuitive to help find repeated patterns and easy to automate for new security bugs: ``\textit{So any tool which can be very intuitive to automate, to write wrappers around (customer rules) [...]. We start noticing a pattern of some kind, and then we add some piece of code to automate that process [...] kind of just taking chunks out of the manual work you have to do.}''.


\cooltitle{Non-functional requirements in security tools \limitation{}:} In our interviews, participants seemed unhappy with the way tools are implemented (P3, P5, P6, P8, P9, P16). Participants reported the concerns related to the base framework (P5), scalability (P8, P16), reliability (P6), efficiency or execution time, and run-time issues (P9, P3). P3 seemed unhappy with the way tooling is built by utilizing multiple frameworks. Similarly, P9 questioned the trustworthiness of Sonarqube because of runtime issues. P8 pointed out scalability as one of their biggest concerns. A tool may fulfill the functional requirement, but it will only be useful when it can scale well as the code base grows. 

\subsection{Learning and Resources (RQ4)}
\label{lab:learningsecurity}

Learning and improving security knowledge is recurrent need among software teams. This section discusses the different learning activities and resources that software practitioners and their team have been using to keep up-to-date with the latest security practices. Table~\ref{tab:learningandresources} shows the participants who pointed out the resources they use while working on security tasks or preparing for the security.

\begin{table*}[htbp]
  \centering
  \scriptsize
\caption{Learning and Resources}
\begin{tabular}{lp{25em}p{10em}}
\toprule
\textbf{ID} & \textbf{Learning strategy or resource} & \textbf{Participants} \\
\midrule
\multicolumn{3}{l}{\textbf{Security tasks}}\\
R1  & Local team and colleagues & P5,P6,P8,P10,P12,P16 \\
R2   & Google Search &  P3, P5, P8, P10, P12\\
R3   & Documentation and Knowledge bases & P3, P6, P8, P13 \\
R4   & OWASP &  P10, P13, P16\\
R5   & StackOverflow &  P10\\
R6   & SecurityFocus & P16 \\
R7   & Programming language crash courses & P16 \\
R8   & Security blogs & P5 \\
\hdashline
\multicolumn{3}{l}{\textbf{Readiness}}\\
R9   & Security training & P1-P3,P6-P8,P10-P15 \\
R10   & Professional certifications & P1,P2,P6,P7,P8,P11,P15 \\
R11   & Capture the flag (CTF) events. & P2, P6, P8, P10, P11 \\
R12   & Self-learning (e.g. setting up a lab) & P1, P14, P16\\
R13   & Security news & P2\\
R14   & Security awareness campaign & P1\\
R15   & Books/articles & P2, P5, P6\\
R16   & Security conferences & P11\\

\bottomrule

\end{tabular}%
\label{tab:learningandresources}

\end{table*}


The \textbf{first order of help for security-related issues or tasks is the local team and colleagues} \resource{} (P5, P6, P8, P10, P12, P16). This makes collaboration (SP22, SP23, SP24 in Section~\ref{lab:collaboration}) between the colleagues a must for resolving the security issues faster. P12 exemplifies: ``\textit{First, of course, I talk to my colleagues. See if somebody has the knowledge. If that fails, I Google like everybody else}''. 

\textbf{Freely searching on the internet} is also a popular practice among practitioners.
Apart from Google Search (P3, P5, P8, P10, P12) \resource{}, participants point out the use of internal and external documentation and knowledge bases (P3, P6, P8, P13) \resource{}, OWASP (P16) \resource{}, stackoverflow\footnote{https://stackoverflow.com/} (P10) \resource{}, the securityfocus website\footnote{https://securityfocus.com} (P16) \resource{}, programming languages crash courses for secure code review (P16) \resource{}, and security blogs (P5) \resource{}.


\textbf{Security training} \resource{} (P1-P3, P6-P8, P10-P15) and \textbf{professional certifications} \resource{} (P1, P2, P6, P7, P8, P11, P15) \textbf{are among the most preferred methods for readiness}. Participants named the SANS\footnote{https://www.sans.org/}, Udemy\footnote{https://www.udemy.com/}, and Pluralsight\footnote{https://www.pluralsight.com/} online training platforms (P1, P2, P3, P8, P11), the CASS, AWS, ISC, and CISSP professional certifications, and the OWASP (discussed in SP7 in Section~\ref{lab:securityguidelines}) as highly recommended online resources for learning application security. Based on the role and responsibilities of the software practitioners, these choices of training and certification differ. P2 explains: ``\textit{There are good courses and organizations that hand out great certificates, one of them is (ISC). They have good certifications, e.g., CISSP. [...] Nowadays, the manufacturers permit these certifications, e.g., an AWS specialist is not only an AWS specialist about the components and functionalities anymore, but also on how to make secure architectures, how to secure the components in this cloud/environment}''. Participants also mentioned that their organizations have a budget for training and often buy enterprise level (or multi-user) subscriptions of these online training resources (P1-P3, P7, P8, P10, P11, P15). P8 explains: ``\textit{[...] we have the personal budget, and you define it together with your manager, where you actually would like to take, for example, offensive security training or e-learn security training, whatever training that you would like to follow and you just get a budget approval [...]}''. 

Apart from online training, the participants pointed out the internal and external classroom training
for preparation of security in advance, where experienced security experts deliver the internal or external training (P1, P6-P8, P10-P15). P14 pointed out that security training is provided within the company, and one should be aware of what kind of training is required to keep oneself updated. One of the participants (P6) pointed out that you can always request a demonstration or presentation from colleagues (SP22, SP23 in Section~\ref{lab:collaboration}. He explains: ``\textit{You always have the option of requesting a presentation from your teammates if we feel that one team member has some knowledge in some area which the rest of the team lacks. [...] We can just ask him to give a presentation, then to keep us in the loop.}''

The \textbf{capture-the-flag (CTF) security events are also a valuable resource to build an advanced security skill-set} \resource{} (P2, P6, P8, P10, P11). A CTF event generally includes a series of challenges of different degrees and requires the participants to apply different skill-sets to solve. These events, along with some training such as OSCE\footnote{https://help.offensive-security.com/hc/en-us/articles/360046801331-OSCE-Exam-Guide}, OSCP\footnote{https://help.offensive-security.com/hc/en-us/articles/360040165632-OSCP-Exam-Guide} are to improve the skill-set for advanced levels of security. P11 explains: ``\textit{If you want to be an advanced penetration tester, you need to make sure that you are sending those penetration testers to those kinds of training, making sure they have enough time to participate in capture the flag in tournaments or do OSCP or OSCE kind of certifications.}''. 

Finally, participants also affirm that \textbf{self-learning (e.g., setting up a lab) \resource{} (P1, P14, P16), security news \resource{} (P2), security awareness campaigns \resource{} (P1), books/articles \resource{} (P2, P5, P6), and security conferences \resource{} (P11)} are also good resources to keep the software practitioners informed and updated on new developments in the field of software security.
For example, P1 mentioned that his company runs the awareness campaign, ``\textit{We have different ways. We have an awareness campaign. Also, virtually, we send emails with information.}'' and P2 pointed out that security news can help prepare for security, ``\textit{What I always do, is read the most recent security news. [...] I think that the literature gives you many ideas and provides you with a lot of information, which will be useful. [...] And this information can be found at anonymous sources. Nowadays, there is MISP\footnote{https://www.misp-project.org/} (Malware Information Sharing Platform).}''.

\subsection{Challenges in Applying Security (RQ5)}
\label{lab:challenges}

The variety of security considerations in different phases of SDLC brings about challenges in applying security. In our interviews, participants provided insight into the challenges encountered while applying security to their current project. In the following, we list (Table~\ref{tab:securitychallenges}) the main security challenges that software development teams face in financial services. 

\begin{table*}[htbp]
  \centering
  \scriptsize
\caption{Challenges in applying security}
\begin{tabular}{lp{25em}p{10em}}
\toprule
\textbf{ID} & \textbf{Challenge} & \textbf{Participants} \\
\midrule
Ch1  & The proliferation of third-party libraries. & P6, P12, P16 \\
Ch2   & Handling blocker security vulnerabilities (CVEs). &  P6, P7, P10-P13\\
Ch3   & Agile methods are too fast to incorporate security. & P4, P6, P15, P16 \\
Ch4   & Organizations' culture and attitude towards security. &  P4, P13, P15\\
Ch5  & Changing requirements. &  P7, P10, P12, P1\\
Ch6  & Following the security guidelines. &  P4,P6,P8,P13,P15,P16\\
Ch7   & Release-time security workload. & P1,  P6,  P8 \\
\bottomrule

\end{tabular}%
\label{tab:securitychallenges}

\end{table*}
\cooltitle{Proliferation of third-party libraries and their management \challenge{}: }
\label{lab:thirdpartylibraries}
In our interviews, participants P6, P12, P16 point out the security challenges with third-party libraries. We infer from the data that the core of this challenge is developers pulling and selecting the libraries (sometimes multiple versions of the same library) based on their requirements to deliver features faster (O12 in Section~\ref{lab:Implementation}). The proliferation of third-party libraries may lead to outdated libraries and build time issues, which further causes the delay in product release and patch management. 
Participant P12 explains: ``\textit{One of the thorniest problems that I have encountered both at a previous company and current is the proliferation of libraries that developers pull-in"}.

Moreover, without a careful selection of third-party libraries, libraries that contain vulnerabilities might be inserted in the product. As P16 observed, while there are tools (known as software composition analysis tools) that can help in detecting such vulnerable dependencies, such tools are helpful just with detection; they do not help developers with how the vulnerable version is being used, and investigating it involves manual work. P16 explains: ``\textit{The software composition analysis tools, the one you use to scan dependencies will give you zero relevant alerts. They'll tell you this dependency might have a vulnerability, but they have no idea how you're using it in your application, so they can't tell you if there is or is not an attack vector due to this dependency.''}

\cooltitle{Blocker CVEs \challenge{}:}
\label{lab:blockercves}
Participants P6, P7, P10-P13 point out the blocker CVEs as another major challenge. A security vulnerability is called a blocker CVE if it blocks any development activity. Blocker CVEs may result in blocking product releases, changes in individual and team's priorities and functioning, and even changes in the product's overall architecture.

Participants P6, P10, P11, P13 emphasize the change in priorities due to blocker CVEs. According to them, the security team needs to make sure that a blocker CVE becomes the developers priority, even if it requires stopping the feature development. P6, a security analyst explains: ``\textit{[...] given the sensitivity of the areas where that CVE is occurring, we can say [...] as a developer, you will fail to proceed. You can no longer proceed with your work until this is 100\% fixed. This and this can, of course, hurt productivity a little bit, but it's a necessary evil.}"

\cooltitle{Agile methods are too fast to incorporate security \challenge{}:}
\label{label:competingwithagilemethod}
Participants P4, P6, P15, P16 point out that agile methodologies might be a roadblock in applying security. According to them, the developers are trained for developing features fast, and the whole idea of agile methods is focused on faster delivery. A two-week sprint may deliver the feature; however, incorporating security aspects may ``need more than one sprint''. 
P15 mentioned this cultural aspect where security is seen as an external entity and developers have difficulty in finding value in what the security team does. According to P4 the decision-making in agile methods goes too fast to incorporate security. P4 says: ``\textit{This part I have identified in the last three years, especially in the financial sector, we are working on an agile security method. So we were already working with agile methodology when developing. But security was not as fast as this agile way of working.}".

\cooltitle{Organizations' culture and attitude towards security \challenge{}: }
\label{lab:resistanceforchange}
The organization's culture and priorities may become a challenge in applying security (P4, P13, P15). According to P15, one of the challenges in applying security is the knowledge gap between the roles. For example, the product owner's decision to overrule threat modeling can create a debatable situation with security practitioners, and this is how P15 describes it as a day-to-day job. Therefore, raising awareness about security should be a priority. P4 says: ``\textit{There are experts, who know how to ensure security. But if there isn't this cultural change within the company, this process doesn't work, or only partially works. I think this is the biggest obstacle.}" 

The security awareness and attitude towards security changes overtime. P13 explains: ``\textit{[...] the lack of security awareness initially; however, it eventually changes based on the team's attitude.''} 
The team's attitude towards security guidelines and incorporating security from the beginning can help avoid such operational challenges later.

\cooltitle{Changing requirements \challenge{}: }
\label{lab:customerrequirements}
The constant changes in requirements are another challenge for the security practitioners, as they need to ensure that security concerns are also addressed (P7, P10, P12, P16). Participants pointed out that the changing requirements can be in the form of new customer scenarios (therefore, new requirements), scenarios not addressed during planning and design, regression testing, late reactions from stakeholders, the architectural issues because of data flow, etc.

Architectural issues often have more priority over other types of security issues and such issues can generate new product requirements (P10, P12). For example, a major change of architecture/design (e.g., saving personally identifiable information) requires encryption (a new requirement) and in-memory data management (while keeping the processing time as short as possible -- yet another new requirement).

Sometimes the customers' expectations may require organizations to demonstrate a higher standard for security assurance as a provider. P12 pointed out that some of their customers are big companies with their own security assurance policies. This demands incorporating the new customer requirements based on their security policies. P12 explains: ``\textit{[...] you have to realize our customers are always big corporate, so it's always large. International banks, petrochemical companies, pharmaceuticals. It's always complicated by the companies [...] (because) they have very stringent security policies, and we have to show that we comply with them.}''

\cooltitle{Following security guidelines \challenge{}: }
\label{lab:followingsecurityguidelines}
Security guidelines often exist (SP10 in Section~\ref{lab:securityguidelines}), but are not followed properly or are too ambiguous to follow (P4, P6, P8, P13, P15, P16). For example, developers overlooking the latest versions of third-party libraries they use or not knowing about existing vulnerabilities in libraries they use, is a matter of failing to comply with security guidelines (P6, P8). Participant P6 pointed out the challenges related to following the security guidelines while working with developers: ``\textit{[...] why would the developers do such things (not following the guidelines)? And why would they be using such vulnerable pieces of code and outdated dependencies? [...] the security parts can become just some green checks you have to get on a pipeline and not something you actively try to achieve or something that you actively try to get done.}''
Moreover, P4 pointed out the scenario where the security guidelines exist, but become unusable. He explains: ``\textit{[...] there is a security test of type X, but there is no adequate order in which they are executed because we didn't establish one in the guidelines. [...] they are established, but on a higher level, and therefore ambiguous. So, in the end, if I have activities and vague or ambiguous guidelines, I don’t apply any order to the activities. It is practically as if we didn't have the guidelines in the first place.}''. 

\cooltitle{Release-time workload \challenge{}: }
\label{lab:releasetimeworkload}
Participants P1, P6, P8 point out the increase in workload at release time, which also changes the existing priorities. The security team with limited resources has work items from multiple projects, and a sudden burst in work items from one or more projects due to approaching release-time affects other projects. P6 explains: ``\textit{For instance, if it happens that several teams have a release, we need to process many items on their end and it can disrupt. It can disrupt a little bit the rest of the projects we are working on.}". Participant P1 even affirms that these disruptions may create trust issues between the developers and the security experts because workload may not allow security engineers to address developer's security related tasks.

\section{Discussion}
\label{label:discussion}

\subsection{Recommendation to Practitioners}

\vspace{2mm} \textbf{Need to understand positioning of security in agile processes.} We observed a mismatch between the way agile processes work and the need for security. Most of the modern software development is agile in nature. Incorporating software security requirements in a traditional agile process may not be as natural (Ch3 and Ch4 in Section~\ref{lab:challenges}).

Other researchers have looked into the ways to fit security assurance into agile methods or vice-versa. The meeting of software assurance and agile development is yet to happen. Earlier work on agile security assurance explored the large gap between security assurance methods and agile methods~\cite{10.1145/1065907.1066034,6046004}. For a long time, the work on integrating agile methods and security assurance has been theoretical~\cite{10.1145/1852786.1852860}. The proposed solutions of bending agile methods to integrate security such as feature driven development (FDD)~\cite{1385609} has security limitations mainly because security is treated as a non-functional requirement~\cite{FIRDAUS2014546}. Other approaches~\cite{6702438,10.1145/1065907.1066034} of including security assurance methods into agile development methods result in additional workload for developers. 

The delegation of security to the security team (Section~\ref{lab:securityasaservice}) and collaboration (Section~\ref{lab:collaboration}) does fit the agile template but not without causing roadblocks for the software development (Ch2 and Ch6 in Section~\ref{lab:challenges}). We argue that more research is required to identify when and how to incorporate secure software engineering practices in agile processes without causing roadblocks for development. Based on our findings, our current recommendation for software teams is to reserve an appropriate amount of time to identify the security requirements in all the development phases and to address the security backlog with priority.

\vspace{2mm} \textbf{Collaboration and team dynamics are key in ensuring security.}
The compliance requirements and maintaining a good security posture are the priority in financial services companies (SP1 in Section~\ref{lab:securityposture}). Our results show that ensuring security demands a high degree of collaboration (S3, S4, S5, S7, S8, S18, S20, S21 in Section~\ref{lab:securityinsdlc}, SP2-SP4, SP23-SP24 in Section~\ref{lab:ensuringsecurity}, R1, R10 in Section~\ref{lab:learningsecurity}). The amount of time spent from reporting security incidents up to their fix is large, and it requires collaboration from experts of different domains. Moreover, during the software system development, a good amount of collaboration, especially between the development and security teams, happens. Oueslati et al. conducted the literature review to identify the challenges of secure development using agile methods and identified ``Collaboration'' as one of the challenges ~\cite{7299963}. On the same line, Kanniah et al. identified the ``Collaboration between the security experts, developers and other stakeholders'' as one of the factors affecting the secure software development practices in the literature review~\cite{kanniah2016review}. On the positive side, the need for collaboration matches with the agile manifesto~\cite{Manifest25:online} which states that collaboration should be prioritized. Collaboration and team dynamics are equally important as the technical aspects of software security. We believe that organizations investing in collaborative infrastructure and team dynamics (for managing security incidents) are likely to benefit in solving security-related issues faster and maintaining good security posture.

\vspace{2mm} \textbf{Awareness of security concerns is growing across the spectrum.}
On a positive note, our interviewees pointed out that the overall state of security has improved in the last decade, as far as financial services companies are concerned. According to them, customers are more aware of the security concerns, and organizations understand that security is a continuous process. Additionally, the end-user should understand the difference between security technologies and secure development. A lack of understanding can give a false notion of security, i.e., ``\textit{putting the application behind the firewall}'' or the attitude of ``\textit{I don't need more security because my application has never been attacked in the past, so it is secure}''. We nevertheless argue that building such awareness (SP2 in Section~\ref{lab:organizationalsecurityframework}) among all the stakeholders should be seen as an on-going process that never stops.

\subsection{Recommendation to Researchers and Educators}

\vspace{2mm} \textbf{Tool inaccuracy remains one of the biggest challenges.} As we observed, security assurance in financial services is highly dependent on automated tools (O13 in Section~\ref{lab:Implementation}, O15 in Section~\ref{lab:testing}). However, our results show that such tools are still of limited use and accuracy (L1, L2, L4 in Section~\ref{lab:securitytoolslimitations}). For example, one of the participants (P16), who has 24 years of professional experience, pointed out that tools are highly inaccurate. This is the reason human intervention is required for code review. 

The software engineering community has been investing in software security research, and this paper only reinforces the need for such investments. We suggest researchers to partner up with financial services companies as they are in clear need of such tools. Finally, there is a lot of investment from financial services companies in the tool section process because a major part of software assurance depends on automated tools. Collaborating on a common platform or being part of an initiative such as OpenSSF\footnote{https://openssf.org/} can benefit all, especially from security tooling.

\vspace{2mm} \textbf{Security training must be a continuous process} Our results show that ``learning security on-the-job'' is a common practice for software engineers (R1-R9 in Section~\ref{lab:learningsecurity}). We have seen companies that have employees dedicated to train other developers on security-related skills (R9, R11 in Section~\ref{lab:learningsecurity}); other companies have dedicated security champions whose goal is to bridge the gap between the security and the development team (SP3 in Section~\ref{lab:organizationalsecurityframework}). 

Gasiba et al.~\cite{gasiba2021secure} conducted a survey to identify the developers' awareness and compliance to secure coding guidelines. The result of the survey is a set of 11 action items for the practitioners focusing on general issues (e.g. management collaboration, security awareness) and secure coding guidelines (SCG) mostly referring to practitioners' lack of security knowledge (e.g. keeping up-to-date with latest technology, not to use SAST in replacement of SCG). We see this as a clear opportunity for researchers and educators to develop and deliver training based on the identified industry requirements.

\section{Threats to validity}
\label{label:ttv}

This section discusses the possible threats to validity and the action we took to mitigate them.

\subsection{Threats to Internal Validity} 
In the scope of this paper, threats to internal validity mostly stem from the interview protocol we propose and the qualitative analysis of the semi-structured interviews we performed. 

As we explained in our methodology section, all the authors collectively prepared the interview protocol. We also conducted a mock interview with a developer who had past software development experience, and improved the protocol based on the feedback we got from the developer as well as our own observations of how the interview went. In practice, we observed participants being very willing to share their current challenges and practices. Therefore, we have no reason to believe that our interview protocol led participants to focus more on one topic or another or to talk, or to favor one specific practice over another.

Most of the qualitative analysis was performed by the first author of this paper. We acknowledge that the analysis may reflect, even unconsciously, the first author's views on software security. For transparency, we make the readers aware that the first author is an expert in the area of software security and had ten years of industry experience at Microsoft India. To mitigate this bias as much as possible the analysis was reviewed by the second and third authors of this paper. In addition, we make our code book available in our online appendix~\cite{appendix_online}.

\subsection{Threats to External Validity}

In the case of this research, threats to external are mostly related to whether our results generalize to other financial services companies.

The participants in our study come from varied financial companies, spread over three continents but mostly in Europe and South America. Most of the companies in our dataset have more than 250 employees, although our sample also contains small and medium companies. The participants themselves are also highly experienced (4 to 32 years) and experts in their profession. 

Nevertheless, we see two possible threats. First, participants were selected via convenience sampling (i.e., via the professional network of the authors). In other words, this sample may be biased towards specific views in software security in ways we can not predict. 
After 16 interviews, authors agreed that saturation of information was achieved. Nevertheless, 
we do not argue that our findings are representative of all types of financial services companies out there, and more replications are needed. We explicitly describe the sample of participants and companies we interviewed (see Table~\ref{tab:participants}), hoping that readers will be able to establish relationships between the context we study here and the context they are inserted into.

\section{Conclusion}
\label{label:Conclusion}

Ensuring that software systems are secure is a challenging and yet fundamental part of modern software engineering. Understanding how practitioners have been handling security concerns in the wild is a stepping stone towards improving the state-of-practice. 
This paper describes the contemporary security considerations and practices in the financial services industry after interviewing 16 software practitioners from 11 financial services companies in 3 continents. 

Our results show how security is integrated into the different stages of the software development life cycle (S1-S23), the security practices that practitioners apply to ensure the security of the software systems they build (SP1-SP23), the key limitations of today's security tools (L1-L5), the different internal initiatives and resources that practitioners use to learn, share, and address security issues (R1-R16), and finally the current challenges they face in applying security (Ch1-Ch7).

Our results emphasize the need to rethink the role of security in agile processes, encourage collaboration and team dynamics to handle security well. Our results also indicate that security tools remain crucial, yet today's tools' accuracy is insufficient. Lastly, as security is a field where systems are under constant attack, continuous learning and training must be a top priority for both industry and academia.

We hope this paper will bring the software development community one step further in securing their software systems. We believe that any software development company can learn from the security initiatives we observed in the financial services.

\bibliographystyle{abbrv}
\bibliography{refs}











\end{document}